\documentclass[12pt]{article}
\usepackage{amssymb,amsmath,epsfig}
\allowdisplaybreaks

\begin{document}
\title{\bf Instability of Meridional Axial System in $f(R)$ Gravity}
\author{M. Sharif \thanks{msharif.math@pu.edu.pk} and Z. Yousaf
\thanks{zeeshan.math@pu.edu.pk}\\
Department of Mathematics, University of the Punjab,\\
Quaid-e-Azam Campus, Lahore-54590, Pakistan.}

\date{}

\maketitle
\begin{abstract}
We analyze dynamical instability of non-static reflection axial
stellar structure by taking into account generalized Euler's
equation in metric $f(R)$ gravity. Such an equation is obtained by
contracting Bianchi identities of usual anisotropic and effective
stress-energy tensors, which after using radial perturbation
technique gives modified collapse equation. In the realm of
$R+\epsilon R^n$ gravity model, we investigate instability
constraints at Newtonian and post-Newtonian approximations. We find
that instability of meridional axial self-gravitating system depends
upon static profile of structure coefficients while $f(R)$ extra
curvature terms induce stability to the evolving celestial body.
\end{abstract}

\section{Introduction}

General relativity (GR) is believed as a remarkable effort in
mathematical physics to analyze gravitational effects of stellar
relativistic interiors. Several interesting consequences coming from
cosmic microwave background, observational ingredients of Supernovae
Ia and its cross-juxtaposition with foreground stellar galactic
distributions \cite{e1}-\cite{e3} have made revolution thereby
introducing new research window. In this realm, many astrophysicists
found GR modifications as helpful to explore unknown aspects of
cosmic gravitational dynamics. The $f(R)$ gravity \cite{e4} is among
extended gravity theories obtained by replacing Ricci invariant with
its generic function $f(R)$ in the Einstein-Hilbert action.

Anisotropic effects are leading paradigms in addressing the
evolutionary mechanisms of celestial imploding models. Herrera and
Santos \cite{e8} reviewed contributions of locally anisotropic fluid
arrangements on the dynamical phases of collapsing shear and
shear-free compact objects. Di Prisco \textit{et al.} \cite{e9}
investigated dynamical properties of anisotropic spherical matter
distribution and found that little fluctuations of pressure
anisotropy lead to system cracking. Sharif and his collaborators
\cite{e11} analyzed effects of anisotropy on the dynamical
properties of spherical as well as non-spherical dense relativistic
distributions and found much complicated system phases due to the
presence of anisotropy. Sunzu \textit{et al.} \cite{e13} studied
analytical models of spherical anisotropic interiors and found that
anisotropic effects provide a broader platform to discuss various
forms of stellar relativistic systems. Recently, we \cite{123}
explored the dynamical features of anisotropic relativistic
interiors.

The spinning stellar distributions indicate direct relevance of
anisotropy with gravitational evolution in which gravitational
radiations cause vorticity within observer congruences. Vorticity
represents rotation of neighboring fluid about an observer moving
with relativistic matter distributions relative to an inertial
frame. Herrera \textit{et al.} \cite{e14} argued that such vorticity
seeds from the existence of super-energy flow which may have direct
relevance with super-Poynting vector. Bonnor \cite{e15} found
electromagnetic energy flow in a relativistic compact distribution
by formulating a relationship between super-Poynting vector and
vorticity. Korunur \textit{et al.} \cite{e16} calculated various
kinematical variables like angular momentum, energy and momentum of
matter configurations associated with an axially symmetric scalar
field. Li \cite{e17} explored superradiant instability of rotating
compact relativistic objects in higher dimensional theory and found
unstable configurations against scalar field perturbations.
Recently, Herrera \textit{et al.} \cite{e18} presented a formal
analysis of gravitational radiations within anisotropic non-static
reflection axial symmetric source and existence of super-energy flow
linked with matter vorticity.

Stability analysis of self-gravitating stellar systems in GR as well
as modified gravity have attracted many researchers for last few
years. The study of different collapsing celestial models with extra
degrees of freedom has great significance to explore late-time
cosmological evolution. Chandrasekhar \cite{e19} discussed
instability constraints for spherical symmetric relativistic
geometry coupled with ideal matter configurations using ratio of
specific heats known as stiffness parameter, $\Gamma_1$. Herrera
{\it et al.} \cite{e20} investigated stability regions for radiating
collapsing stellar objects and concluded that dissipation vector
tends to move the systems towards stable configurations. Chan {\it
et al.} \cite{e21} studied remarkable effects of shearing viscosity
and anisotropy on the instability constraints at Newtonian (N) and
post-Newtonian (pN) eras.

Cai \cite{e22} discussed dynamical properties and structure
formation of dense matter relativistic configurations in modified
gravity by assigning zero, negative, and positive values of constant
curvature. Bamba \textit{et al.} \cite{bamba1} performed dynamical
analysis of a collapsing relativistic stellar system and claimed
that invoking of $R^\alpha(1<\alpha\leq2)$ corrections could helps
to present a viable and singularity free model. Myung \textit{et
al.} \cite{myung1} performed stability analysis of sphericalstellar
structure with constant Ricci invariant background in metric $f(R)$
gravity via perturbation scheme and noticed relatively stable
distributions under specific constraints. Moon \textit{et al.}
\cite{moon1} extended these consequences with negative cosmological
constant environment and calculated limits for the stability of
relativistic systems.

Capozziello \textit{et al.} \cite{e23} explored dynamical evolution
of relativistic collapsing spherical interior in $f(R)$ gravity by
evaluating extended form of Poisson and Boltzmann equations. De
Laurentis and Capozziello \cite{e24} discussed instability issue of
stellar interior at N approximation with $f(R)$ extra degrees of
freedom and also studied axisymmetric black hole models. Astashenok
\textit{et al.} \cite{e25} investigated evolution of
self-gravitating systems and found relatively more massive and
supergiant dense configurations due to $f(R)$ gravity corrections.
Farinelli \textit{et al.} \cite{e26} discussed dynamical properties
of stellar systems in the presence of $f(R)$ corrections and found
that higher degree terms tend to mollify collapsing process. Sharif
and his collaborators \cite{e27,e28} studied instability constraints
for restricted class of axially symmetric spacetime by means of
adiabatic index/stiffness parameter.

The present paper aims to extend our previous work \cite{e28} of
stability analysis by taking reflection effects in non-static axial
symmetric anisotropic source with $\epsilon R^n$ extra degrees of
freedom. In the present paper, we develop instability regions for
anisotropic meridional axisymmetric source with $R+\epsilon R^n$
background. The inclusion of $\epsilon R^n$ correction in our
analysis seeds from the fact that this correspond to the various
eras of the cosmic history thereby helping to explain the
gravitational dynamics during inflationary as well as late-time
accelerating regimes. Furthermore, the addition of meridional
effects in stellar system causes a flow of gravitational energy due
to existence of vorticity tensor in the analysis.

The paper has the following format. Section $\textbf{2}$ deals with
kinematical formulations of comoving meridional axial symmetric
geometry coupled with anisotropic matter configurations. The
meridional effects in stellar system causes a flow of gravitational
energy due to existence of vorticity tensor. We present $f(R)$ dark
source components and set of dynamical equations with reflection
axial degrees of freedom. In section $\textbf{3}$, we discuss viable
$f(R)$ model and use perturbation method to develop collapse
equation. Section \textbf{4} explores instability constraints.
Finally, we summarize our results in the last section.

\section{Anisotropic Source and Field Equations}

The extended configuration of Einstein-Hilbert action is
\begin{equation}\label{1}
S_{f(R)}=\frac{1}{2\kappa}\int d^4x\sqrt{-g}f(R)+S_M,
\end{equation}
where $\kappa,~f(R),~S_M,~T_{\alpha\beta}$ are coupling constant,
matter action, a non-linear Ricci curvature function and usual
stress energy tensor, respectively. The variation of above action
with respect to $g_{\alpha\beta}$ provides the field equations
\begin{equation}\label{2}
f_RR_{\alpha\beta}-\nabla_{\alpha}
\nabla_{\beta}f_R-g_{\alpha\beta}\left(\frac{1}{2}f-{\Box}f_R\right)={\kappa}T_{\alpha\beta},
\end{equation}
where $\Box,~\nabla_\alpha,~f_R$ are D'Alembert, covariant
derivative and $\frac{df}{dR}$ operators, respectively. Equation
(\ref{2}) can be written in terms of Einstein tensor as
\begin{equation}\label{3}
G_{\alpha\beta}=\frac{\kappa}{f_R}(\overset{(D)}
{T_{\alpha\beta}}+T_{\alpha\beta}),
\end{equation}
where
\begin{equation}\label{3a}
\overset{(D)}{T_{\alpha\beta}}=\frac{1}{\kappa}\left\{\frac{R}{2}\left(\frac{f}{R}-f_R\right)
g_{\alpha\beta}-{\Box}f_Rg_{\alpha\beta}+\nabla_{\alpha}\nabla_{\beta}f_R\right\},
\end{equation}
is the stress energy tensor indicating $f(R)$ contribution in the
dynamics of relativistic systems. We take axially symmetric metric
characterizing reflection effects \cite{e18}
\begin{equation}\label{4}
ds^2=-A^2(t,r,\theta)dt^{2}+2L(t,r,\theta)dtd\theta+B^2(t,r,\theta)(dr^{2}+r^2d\theta^{2})
+C^2(t,r,\theta){d\phi^2},
\end{equation}
with locally anisotropic fluid configuration
\begin{align}\nonumber
T_{\alpha\beta}&=(\mu+P)V_{\alpha}V_{\beta}+Pg_{\alpha\beta}
+\frac{1}{3}(\Pi_{II}+2\Pi_I)(K_\alpha
K_\beta-\frac{1}{3}h_{\alpha\beta})+\frac{1}{3}(\Pi_I+2\Pi_{II})\\\label{5}
&\times(N_{\alpha}N_\beta-\frac{1}{3}h_{\alpha\beta})+\Pi_{KN}(K_{\alpha}N_\beta+K_\beta
N_\alpha),
\end{align}
where $\mu,~P,~\Pi_I,~\Pi_{II},~\Pi_{KL}$ and $h_{\alpha\beta}$ are
the fluid energy density, pressure, anisotropic scalars and
projection tensor, respectively. The matter four velocity,
$V_\alpha$, and spacelike vectors $S_\alpha,~K_\alpha$ and
$N_\alpha$ in comoving coordinates are
\begin{equation}\label{6}
V^\alpha=\frac{1}{A}\delta^0_\alpha,~V_\alpha=-A\delta^0_\alpha
+\frac{L}{A}\delta^2_\alpha,~S_\alpha=C\delta^3_\alpha,~K_\alpha
=B\delta^1_\alpha,~N_\alpha=\frac{\sqrt{\Delta}}{A}\delta^2_\alpha,
\end{equation}
where $\Delta=(ABr)^2+L^2$, which obey the following constraints
\begin{align}\nonumber
&K^\alpha N_\alpha=K^\alpha S_\alpha=S^\alpha N_\alpha=V_\alpha
K^\alpha = V^\alpha N_\alpha = V^\alpha S_\alpha=0,\\\nonumber
&K_\alpha K^\alpha=N_\alpha N^\alpha=S_\alpha S^\alpha=-V^\alpha
V_\alpha=1.
\end{align}
The fluid pressure and its anisotropic scalars can be expressed
alternatively in terms of projection tensor and spacelike vectors,
respectively as
\begin{align}\nonumber
&P= \frac{1}{3}h^{\alpha\beta}T_{\alpha\beta},\quad
\Pi_{I}=(2K^\alpha K^\beta -S^\alpha S^\beta-N^\alpha
N^\beta)T_{\alpha\beta},\quad
\Pi_{KN}=K^{\alpha}N^{\beta}T_{\alpha\beta},\\\nonumber
&\Pi_{II}=(2N^{\alpha}N^\beta-K^\alpha K^\beta-S^\alpha
S^\beta)T_{\alpha\beta}.
\end{align}

The non-zero components of effective stress energy tensor (\ref{3a})
are
\begin{align}
&\overset{(D)}{T^{\alpha\beta}}=\left[ {\begin{matrix}\label{123}
V_1+W_1&X_1+Y_1&X_3+Y_3&0 \\
X_1+Y_1&V_2+W_2&X_2+Y_2&0 \\
X_3+Y_3&X_2+Y_2&V_3+W_3&0 \\
0&0&0&V_4+W_4 \\
\end{matrix} } \right],
\end{align}
where dark source $f(R)$ terms $V_i,~W_i$ and $X_j,~Y_j$ are
diagonal and non-diagonal components of effective energy-momentum
tensor (\ref{3a}), respectively, in which $W_i$ and $Y_j$
incorporate axial reflection effects with $f(R)$ extra degrees of
freedom. By choosing $X_j$ and $Y_j$ equal to zero along with
$\Delta\rightarrow{A^2B^2r^2}$, higher curvature terms of restricted
axisymmetric metric can be found. However, the inclusion of these
terms along with anisotropic in usual stress energy tensor ensure
the propagation of gravitational radiations in the environment
\cite{13new}.

The kinematical quantity controlling local spinning motion of
anisotropic matter distributions is the vorticity tensor which for
meridional axially symmetric metric can be expressed in terms of
$K_\alpha$ and $N_\alpha$ as
\begin{equation}\nonumber
\Omega_{\alpha\beta}=\Omega(K_\beta N_\alpha-N_\beta K_\alpha),
\end{equation}
where
\begin{eqnarray}\label{18}
\Omega=\frac{L}{2B\sqrt{\Delta}}\left(\frac{L'}{L}-\frac{2A'}{A}\right)
\end{eqnarray}
is known as vorticity scalar. Here prime stand for
$\frac{\partial}{\partial{r}}$. There exists only one independent
non-zero vorticity component along $r\theta$ direction. The
existence of vorticity scalar is directly related to the existence
of reflection effects of axisymmetric spacetime as it is controlled
by non-diagonal structure coefficient, $L$. Thus if one takes
$\Omega=0$ over the dynamical evolution of axial anisotropic
spacetime, this imparts null value to non-diagonal scale factor
whose dynamics has already been discussed \cite{e28}.

In order to evaluate dynamical evolution equations for axially
symmetric relativistic celestial body with $f(R)$ background, we
consider
\begin{align*}
|T^{\alpha\beta}+\overset{(D)}{T^{\alpha\beta}}|_{;\beta}=0,
\end{align*}
which yields
\begin{align}\nonumber
&\dot{\mu}-\mu\left[\frac{\dot{B}}{B}+\frac{\dot{C}}{C}+\frac{1}
{\Delta}\left(r^2A\dot{A}B^2+L\dot{L}+r^2A^2B\dot{B}\right)\right]+
\frac{AB^2(\mu+P)}{\Delta}\left[r^2\left(\frac{2\dot{B}}{B}\right.\right.\\\nonumber
&\left.\left.+\frac{\dot{C}}{C}\right)+\frac{L^2}{A^2B^2}\left(
\frac{\dot{B}}{B}-\frac{\dot{A}}{A}+\frac{\dot{L}}{L}+\frac{\dot{
C}}{C}\right)\right]+\frac{\Pi_I}{3A}\left(\frac{\dot{B}}{B}-\frac{
\dot{C}}{C}\right)+\frac{\Pi_{II}}{3\Delta}\left\{AB^2r^2\left(
\frac{\dot{B}}{B}\right.\right.\\\label{19}
&\left.\left.-\frac{\dot{C}}{C}\right)+\frac{L^2}{A}\left(\frac{
\dot{L}}{L}-\frac{\dot{A}}{A}-\frac{\dot{C}}{C}\right)\right\}
+\left(\frac{B^2r^2A\dot{A}}{\Delta}+\frac{\dot{C}}{C}\right)
V_1+D_0(t,r,\theta)=0,\\\nonumber
&P'+\frac{2}{9}\left(2\Pi_I'+\Pi_{II}'\right)+\left[P+\frac{2}{9}
\left(2\Pi_I+\Pi_{II}\right)\right]\left[\frac{C'}{C}+\frac{3LL'}
{2\Delta}+\frac{r^2A^2B^2}{\Delta}\left(\frac{A'}{A}\right.\right.\\\nonumber
&\left.\left.+\frac{2B'}{B}+\frac{2}{r}-\frac{(rB)'}{rB}\right)\right]
-\frac{r^2AB^5}{\Delta^{3/2}}\left[\Pi_{KN,\theta}-\left\{\frac{A_\theta}
{A}+\frac{6B_\theta}{B}+\frac{C_\theta}{C}+\frac{4r^2A^2B^2}
{\Delta}\right.\right.\\\nonumber
&\left.\left.\times\left(\frac{A_\theta}{A}+\frac{B_\theta}{B}\right)
+\frac{4LL_\theta}{\Delta}\right\}\Pi_{KL}\right]+\frac{{\mu}r^4A^4B^4}{\Delta^2}
\left(B\dot{B}+\frac{A'}{A}-\frac{LA_\theta}{r^2AB^2}\right)
-\left(\frac{(rB)'}{rB}\right.\\\label{20}
&\left.+\frac{L}{2L'}\right)\frac{{\mu}r^2A^2L^2B^2}{\Delta^2}+\left(
\frac{3\dot{B}}{B}+\frac{r^2B^2A\dot{A}}{\Delta}+\frac{\dot{C}}{C}
\right)X_1+D_1(t,r,\theta)=0,\\\nonumber
&\frac{{\mu}r^2A^2B^2L}{\Delta^2}\left[\frac{\dot{\mu}}{\mu}+\frac{\dot{A}}
{A}+\frac{3\dot{B}}{B}+\frac{\dot{L}}{L}+\frac{\dot{C}}{C}+\frac{1}{r^2B^2}
\left(\frac{\mu_{\theta}}{\mu}+\frac{2L_\theta}{L}+\frac{2A_\theta}{A}\right)
+\frac{1}{\Delta}\left\{4r^2A^2\right.\right.\\\nonumber
&\left.\left.\times\left(\frac{\dot{A}}{A}+\frac{\dot{B}}{B}
\right)-\frac{4\dot{L}}{L}-LA^2\left(\frac{5A_\theta}{A}+\frac{2B_\theta}{B}\right)
+r^2A^2B^2\left(\frac{\dot{L}}{L}+\frac{\dot{B}}{B}\right)+\frac{r^2A^3B^2A_\theta}
{L}\right\}\right.\\\nonumber
&\left.-\frac{4L^2L_\theta\Delta}{r^2B^2}\right]+\frac{\mu
A^2L^2}{\Delta^2}\left\{\frac{B_\theta}{B}+\frac{C_\theta}{C}-\frac{r^2BL
\dot{B}}{\Delta}\right\}-\frac{r^3AB^3\Pi_{KN}}{\Delta^{\frac{3}{2}}}
\left[\frac{\Pi'_{KN}}{\Pi_{KN}}+\frac{3}{r}\right.\\\nonumber
&\left.+\frac{4B'}{B}+\frac{A'}{A}+\frac{C'}{C}+\frac{3}{\Delta}\left
\{\frac{LL'}{2}+r^2A^2B^2\left(\frac{3}{r}+\frac{2A'}{A}+\frac{3B'}{B}\right)
\right\}+\frac{7LL'}{2\Delta}\right]\\\nonumber
&+\frac{1}{\Delta}\left\{P+\frac{2}{9}(\Pi_I+2\Pi_{II})\right\}\left[
\frac{r^2A^2B^2}{\Delta}\left\{(2A^2+A)\left(\frac{A_\theta}{A}+\frac{
B_\theta}{B}\right)-\frac{L\dot{B}}{B}\right.\right.\\\nonumber
&\left.\left.+\frac{2AB_\theta}{B}\right\}+2AA_\theta+\frac{A^2C_\theta}{C}
-\frac{r^2BL\dot{L}}{\Delta}-\frac{2A^2LL_\theta}
{\Delta}-\frac{L\dot{B}}{B}\right]-\frac{P}{C\Delta}(L\dot{C}
\\\label{21}
&+A^2C_\theta)+\frac{A^2}{\Delta}\left\{P_\theta+\frac{2}{9}(\Pi_{I,\theta}
+2\Pi_{II,\theta})\right\}+D_2(t,r,\theta)=0,
\end{align}
where $D_0,~D_1$ and $D_2$ are $f(R)$ corrections given in Appendix
\textbf{A}. Here over dot and subscript $\theta$ stand for
$\frac{\partial}{\partial{t}}$ and
$\frac{\partial}{\partial{\theta}}$, respectively. The second of the
above equations is known as generalized Euler equation.

\section{$f(R)$ Model and Perturbation Scheme}

Many inflationary models in the early universe are established on
scalar fields coming in supergravity and superstring theories. The
first model of inflation was suggested by Starobinsky which deals
with conformal anomaly in quantum gravity \cite{e29} given by
\cite{e30}
\begin{equation}\label{22}
f(R)=R+{\epsilon}R^{n},
\end{equation}
where $n$ can be positive or negative integer. This model explains
the present universe acceleration due to the presence of dark energy
and can serve as power-law inflation, i.e. exponential expansion and
ordinary inflation incorporating minimally coupled scalar field.
Here $\epsilon\sim\frac{1}{M^{2n-2}}>0$ for $n>0$ and $M$ has the
mass dimensions. Since $f(R)$ gravity can be used as an alternative
for dark matter \cite{e31} in addition to dark energy at cluster as
well as stellar scales, so this model with $n=2$ was claimed both as
dark matter model with $\epsilon=\frac{1}{6M^2}$ \cite{e32} and as
dark energy. The value of $M$ is chosen to be $2.7\times10^{-12}GeV$
along with $\epsilon\leq2.3\times10^{22}Ge/V^2$ for dark matter
cosmology \cite{e33}. All GR solutions can be found by taking limit
$f(R)\rightarrow R$.

Here, we use perturbation method \cite{e20,e21} to explore modified
collapse equation for meridional axially symmetric anisotropic
geometry. For very small values of perturbation parameter $\alpha$
with $0<\alpha\ll1$, we take effects up to $O(\alpha)$. We first
suppose that the system is in hydrostatic equilibrium at $t=0$,
however on departing from this state, the system depends upon the
same time dependence factor $T(t)$ on all its structure
coefficients. The structure and matter variables can be perturbed as
follows
\begin{eqnarray}\label{23}
\mathcal{S}(t,r,\theta)&=&\mathcal{S}_0(r,\theta)+{\alpha}T(t)s(r,\theta),\\\label{27}
\mathcal{M}(t,r,\theta)&=&\mathcal{M}_0(r,\theta)+{\alpha}\bar{m}(t,r,\theta),
\end{eqnarray}
where $\mathcal{S}$ represents perturbation method applicable on
structural co-efficients of Eq.(\ref{4}), i.e., $A,~B,~C,~L$ and on
Ricci scalar, $R$ which after perturbation denotes $s$ as
$a,~b,~c,~l$ and $e$, respectively. Equation (\ref{27}) indicates
perturbation method of matter variables (these matter variables are
taken from Eq.(\ref{5})). Thus the allocation of $\mathcal{M}$ will
be $\mu,~P,~\Pi_{a},~a=1,2,3$ and the corresponding perturbed
quantities will be represented by placing bar over that. However,
the perturbation technique for $f(R)$ model is given as follows
\begin{eqnarray}\label{31}
f(t,r)&=&\left[R_0(r)+{\epsilon}R_0^{n})\right]
+{\alpha}T(t)e(r)\left[1-{\epsilon}nR_0^{n-1}\right],\\\label{32}
f_R(t,r)&=&1+{\epsilon}nR_0^{n-1}+{\alpha}T(t)e(r)n{\epsilon}(n-1)R_0^{n-2},
\end{eqnarray}
where $R_0$ represents static distribution of Ricci scalar. Using
Eqs.(\ref{23})-(\ref{32}), the first of dynamical equations
satisfies trivially, while rest of dynamical equations (\ref{20})
and (\ref{21}) at $t=0$ give
\begin{align}\nonumber
&P'_0+\frac{2}{9}\left(2\Pi_{I0}'+\Pi_{II0}'\right)
+\left[P_0+\frac{2}{9}\left(2\Pi_{I0}+\Pi_{II0}\right)
\right]\left[\frac{C'_0}{C_0}+\frac{3L_0L'_0}{2\Delta_0}
+\frac{r^2A^2_0B^2_0}{\Delta_0}\right.\\\nonumber
&\times\left.\left(\frac{A'_0}{A_0}+\frac{1}{r}\right)\right]
-\frac{r^3A_0B^5_0}{\Delta_0^{\frac{3}{2}}}\Pi_{KN0,\theta}
-\frac{r^3A_0B^5_0}{\Delta_0^{\frac{3}{2}}}\left\{\frac{
A_{0\theta}}{A}+\frac{6B_{0\theta}}{B_0}+\frac{C_{0\theta}}{C_0}
+\frac{4L_0L_{_0\theta}} {\Delta_0}\right.\\\nonumber
&\left.+\frac{4r^2A^2_0B^2_0}{\Delta_0}\left(\frac{A_{0\theta}}{A_0}
+\frac{B_{0\theta}}{B_0}\right)\right\}+\frac{\mu
r^4A^4_0B^4_0}{\Delta_0^2}\left(\frac{A'_0}{A_0}-\frac{L_0A_{
0\theta}}{r^2A_0B^2_0}\right)-\frac{\mu_0r^2A^2_0L^2_0B^2_0}
{\Delta_0^2}\\\label{33}
&\left(\frac{L_0}{2L'_0}+\frac{1}{r}+\frac{B_0'}{B_0}\right)+D_{1S}=0,\\\nonumber
&\frac{\mu_0r^2A_0^2B_0^2L_0}{\Delta^2}\left[+\frac{1}{r^2B^2_0}
\left(\frac{\mu_{0\theta}}{\mu_0}+\frac{2L_{0\theta}}{L_0}+\frac{
2A_{0\theta}}{A_0}\right)-\frac{1}{\Delta_0}\left\{L_0A^2_0\left(
\frac{5A_{0\theta}}{A_0}+\frac{2B_{0\theta}}{B_0}\right)
\right.\right.\\\nonumber
&\left.\left.-\frac{r^2A_0^3B_0^2A_{0\theta}}{L_0}\right\}
-\frac{4L_0^2L_{0\theta}\Delta_0}{r^2B^2_0}\right]+\frac{\mu
A_0^2L^2_0}{\Delta^2_0}\left\{\frac{B_{0\theta}}{B_0}+\frac{
C_{\theta0}}{C}\right\}-\frac{r^3A_0B^3_0\Pi_{KN0}}{\Delta_0^{3/2}}
\left[\frac{3}{r}\right.\\\nonumber
&\left.+\frac{\Pi'_{KN0}}{\Pi_{KN0}}+\frac{4B_0'}{B_0}+\frac{A'_0}
{A_0}+\frac{C_0'}{C_0}+\frac{3}{\Delta_0}\left\{\frac{L_0L'_0}{2}
+r^2A_0^2B_0^2\left(\frac{3}{r}+\frac{2A'_0}{A_0}+\frac{3B'_0}{B_0}\right)
\right\}\right.\\\nonumber
&\left.+\frac{7L_0L'_0}{2r\Delta_0}\right]+\frac{1}{\Delta_0}
\left\{P_0+\frac{2}{9}(\Pi_{I0}+2\Pi_{II0})\right\}
\left[\frac{r^2A_0^2B^2_0}{\Delta_0}\left\{(2A^2_0+A_0)\left(
\frac{A_{0\theta}}{A_0}+\frac{B_{0\theta}}{B_0}\right)
\right.\right.\\\nonumber
&\left.\left.+\frac{2A_0B_{0\theta}}{B_0}\right\}+2A_0A_{0\theta}
+\frac{A_0^2C_{0\theta}}{C_0}-\frac{2A_0^2L_0L_{0\theta}}
{\Delta_0}\right]+A_0^2C_{0\theta}+\frac{A_0^2}{\Delta_0}
\left\{P_{0\theta}+\frac{2}{9}(\Pi_{I0\theta}\right.\\\label{34}
&\left.+2\Pi_{II0\theta})\right\}+D_{2S}=0.
\end{align}

The static $f(R)$ contribution of second and third conservation
equations are denoted by $D_{2S}$ and $D_{3S}$, respectively and can
be calculated very easily from Eqs.(\ref{d1}) and (\ref{d2}) after
using perturbation method. Using Eqs.(\ref{23})-(\ref{32}), the
non-static perturbed axial dynamical equation (\ref{19}) will take
the form
\begin{align*}\nonumber
&\dot{\bar{\mu}}+\left[\mu_0\left\{\frac{b}{B_0}+\frac{c}{C_0}+\frac{1}{\Delta_0}
\left(r^2aA_0^2B_0^2+lL_0+r^2bB_0L_0\right)\right\}+(\mu_0+P_0)\frac{A_0^2
B_0^2}{\Delta_0^2}\right.\\\nonumber
&\times\left\{r^2\left(\frac{2b}{B_0}+\frac{2c}{C_0}\right)+\frac{L_0^2}{A_0^2B_0^2}
\left(\frac{b}{B_0}+\frac{l}{L_0}-\frac{a}{A_0}+\frac{c}{C_0}\right)\right\}
+\frac{\Pi_{I0}}{3}\left(\frac{b}{B_0}-\frac{c}{C_0}\right)\\\nonumber
&\left.+\frac{\Pi_{II0}}{3\Delta_0}
\left\{r^2A_0^2B_0^2\left(\frac{b}{B_0}-\frac{c}{C_0}\right)+L_0^2\left(\frac{l}{L_0}-\frac{a}{A_0}
-\frac{c}{C_0}\right)\right\}+D_3(r,\theta)\right]\dot{T}=0,
\end{align*}
where $D_3$ represents $f(R)$ corrections which can be obtained from
expressions $g(t,r,\theta)$ and $h(t,r,\theta)$ given in Appendix
\textbf{A}. Substituting Eq.(\ref{22}) in Eq.(\ref{123}) and then
employing perturbation method, one can obtain $f(R)$ dynamical
quantities, $V_i,~W_i,~X_j,~Y_j$ whose values upon substitution in
Eqs.(\ref{gg}) and (\ref{hh}) yield $D_3$ such that
$g(t,r,\theta)+h(t,r,\theta)=D_3\dot{T}$. Integration of the above
equation gives
\begin{align}\label{35}
&{\bar{\mu}}=-\chi(r,\theta){T},
\end{align}
where \begin{align}\nonumber
\chi&=\left[\mu_0\left\{\frac{b}{B_0}+\frac{c}{C_0}+\frac{1}{\Delta_0}
\left(r^2aA_0^2B_0^2+lL_0+r^2bB_0L_0\right)\right\}+(\mu_0+P_0)\frac{A_0^2
B_0^2}{\Delta_0^2}\right.\\\nonumber
&\times\left\{r^2\left(\frac{2b}{B_0}+\frac{2c}{C_0}\right)+\frac{L_0^2}{A_0^2B_0^2}
\left(\frac{b}{B_0}+\frac{l}{L_0}-\frac{a}{A_0}+\frac{c}{C_0}\right)\right\}
+\frac{\Pi_{I0}}{3}\left(\frac{b}{B_0}-\frac{c}{C_0}\right)\\\nonumber
&\left.+\frac{\Pi_{II0}}{3\Delta_0}
\left\{r^2A_0^2B_0^2\left(\frac{b}{B_0}-\frac{c}{C_0}\right)+L_0^2\left(
\frac{l}{L_0}-\frac{a}{A_0}-\frac{c}{C_0}\right)\right\}+D_3(r,\theta)\right].
\end{align}

Now, we evaluate $t\theta$ component of metric $f(R)$ field
equations (\ref{3}) and then using perturbation scheme along with
some manipulations, it follows that
\begin{eqnarray}\label{36}
\varrho_1\ddot{T}+\varrho_2\dot{T}+\varrho_3T=0,
\end{eqnarray}
where quantities $\varrho_i$ contain combinations of meridional
axial geometric functions as well as $R+\epsilon R^n$ corrections,
depending upon $r$ and $\theta$ coordinates and are assumed
positive. More specifically, these quantities incorporate
non-perturbed as well as perturbed terms. There exist oscillating as
well as non-oscillating solutions of the above equation which
represent unstable as well as stable models of evolving relativistic
stellar systems, respectively. We confine ourselves to obtain
solutions for collapsing relativistic system. Thus we limit our
perturbation parameters, $a,~b,~c,~e$ and $l$ to be positive
definite quantities for which we obtain $\omega^2>0$. In this
context, the solution of Eq.(\ref{36}) is given by
\begin{equation}\label{37}
T(t)=-\exp({\omega}t),\quad\textrm{where}\quad
\omega^2=\frac{-\varrho_2+\sqrt{\varrho_2^2-4\varrho_1\varrho_3}}{2\varrho_1}.
\end{equation}
Using the perturbation technique, the non-static distributions of
Eq.(\ref{20}), after using Eq.(\ref{37}), are written as
\begin{align}\nonumber
&\frac{1}{B_0^2}\left\{\bar{P}'+\frac{2}{9}(2\bar{\Pi}_I'+\bar{
\Pi}_{II}')\right\}+\frac{1}{B_0^2}\left\{\bar{P}+\frac{2}{9}(2
\bar{\Pi}_I+\bar{\Pi}_{II})\right\}\left\{\frac{C_0'}{C_0}
+\frac{3L_0L_0'}{2\Delta_0}+\left(\frac{1}{r}\right.\right.\\\nonumber
&\left.\left.+\frac{A_0'}{A_0}\right)\frac{r^2A_0^2B_0^2}{\Delta_0}
\right\}-\frac{r^3A_0B_0^3}{\Delta^{3/2}_0}\bar{\Pi}_{KN,\theta}
-\bar{\Pi}_{KN}\frac{r^3A_0B_0^3}{\Delta_0^{3/2}}\left\{\frac{A_{0\theta}}{A_0}
+\frac{6B_{0\theta}}{B_0}+\frac{4L_0L_{_0\theta}}
{\Delta_0}\right.\\\nonumber
&\left.+\frac{C_{0\theta}}{C_0}+\frac{4r^2A_0B^2_0}{\Delta_0}
\left(\frac{A_{0\theta}}{A_0}+\frac{B_{0\theta}}{B_0}\right)
\right\}+\frac{\bar{\mu}r^4A_0^4}{\Delta_0^2}
\left(\frac{A_0'}{A_0}-\frac{L_0A_{0\theta}}{r^2A_0B_0^2}\right)
-\bar{\mu}L_0^2r^2\frac{A_0^2}{\Delta_0^2}\\\nonumber
&\times\left(\frac{L'_0}{2L_0}+\frac{B_0'}{B_0}\right)-\frac{2b}
{B_0^3}\left\{{P_0}'+\frac{2}{9}(2{\Pi}_{I0}'+{\Pi}_{
II0}')\right\}T+\frac{r^3A_0B_0^3}{\Delta^{3/2}_0}{\Pi}_{KN0,
\theta}\left(\frac{a}{A_0}\right.\\\label{38}
&\left.+\frac{3b}{B_0}-\frac{3d}{\Delta_0}
\right)T+\frac{lL_0X_{10}}{\Delta_0}+\frac{r^2A_0^2bB_0}{\Delta_0}
X_{30}+[\Upsilon+\Phi+\zeta]T=0,
\end{align}
where $\Upsilon,~\zeta$ and $\Phi$ are mentioned in Appendix
\textbf{A}. The quantity controlling the reflection degrees of
freedom along with $f(R)$ corrections of an axisymmetric celestial
body is $\Upsilon$. However, the expression $\Phi$ incorporates
gravitational contribution due to $f(R)$ gravity, while $\zeta$ is
the remaining part of non-static perturbed generalized Euler
equation holding usual Einstein gravity effects.

In view of second law of thermodynamics, we can link perturbed
anisotropic quantities with energy density by an equation of state
as \cite{e35}
\begin{equation}\label{39}
\bar{P}_{i}=\Gamma_1\frac{P_{i0}}{\mu_0+P_{i0}}\bar{\mu},
\end{equation}
where $\Gamma_1$ is a fluid stiffness parameter also known as
adiabatic index. This measures pressure variations of matter
configurations with respect to energy density. In our analysis,
$\Gamma_1$ will be treated as a constant identity. Using
Eqs.(\ref{35}) and (\ref{39}), we have
\begin{align}\nonumber
\bar{\Pi}_{KN}&=-\Gamma_1\frac{\Pi_{KN0}}{\mu_0+\Pi_{KN0}}\chi
T,\quad \bar{P}=-\Gamma_1\frac{P_{0}}{\mu_0+P_{0}}\chi T,\\\nonumber
\bar{\Pi}_I&=-\Gamma_1\frac{\Pi_{I0}}{\mu_0+\Pi_{I0}}\chi T,\quad
\bar{\Pi}_{II}=-\Gamma_1\frac{\Pi_{II0}}{\mu_0+\Pi_{II0}}\chi T.
\end{align}
Using Eq.(\ref{35}) as well as the above relations in Eq.(\ref{38}),
we obtain
\begin{align}\nonumber
&-\frac{1}{B_0^2}\Gamma_1\phi'T-\frac{1}{B_0^2}\Gamma_1{\phi}T
\left\{\frac{C_0'}{C_0}+\frac{3L_0L_0'}{2\Delta_0}+\left(\frac{
1}{r}+\frac{A_0'}{A_0}\right)\frac{r^2A_0^2B_0^2}{\Delta_0}
\right\}-\frac{r^3A_0B_0^3}{\Delta^{3/2}_0}\Gamma_1T\\\nonumber
&\times\left(\frac{{\Pi}_{KN0}\chi}{\mu_0+{\Pi}_{KN0}}\right)_{\theta}
-\frac{{\Pi}_{KN0}\chi}{\mu_0+{\Pi}_{KN0}}\frac{r^3A_0B_0^3}{
\Delta_0^{3/2}}\left\{\frac{A_{0\theta}}{A_0}+\frac{4r^2A_0B^2_0}
{\Delta_0}\left(\frac{A_{0\theta}}{A_0}+\frac{B_{0\theta}}{B_0}
\right)\right.\\\nonumber
&\left.+\frac{C_{0\theta}}{C_0}+\frac{6B_{0\theta}}{B_0}+\frac{4L_0L_{_0\theta}}
{\Delta_0}\right\}-T\frac{{\chi}r^4A_0^4}{\Delta_0^2}
\left(\frac{A_0'}{A_0}-\frac{L_0A_{0\theta}}{r^2A_0B_0^2}\right)
+T{\chi}L_0^2r^2\frac{A_0^2}{\Delta_0^2}\left(\frac{L'_0}{2L_0}
\right.\\\nonumber
&\left.+\frac{B_0'}{B_0}\right)-\frac{2b}{B_0^3}\left\{{P_0}'
+\frac{2}{9}(2{\Pi}_{I0}'+{\Pi}_{II0}')\right\}T+\frac{r^3A_0
B_0^3}{\Delta^{3/2}_0}\left(\frac{a}{A_0}+\frac{3b}{B_0}
-\frac{3d}{\Delta_0}\right)\\\label{40}
&\times{\Pi}_{KN0,\theta}T+\frac{lL_0X_{10}}{\Delta_0}
+\frac{r^2A_0^2bB_0}{\Delta_0}X_{30}+[\Upsilon+\Phi+\zeta]T=0,
\end{align}
where
$\phi=\frac{P_0\chi}{(\mu_0+P_0)}+\frac{4\Pi_{I0}\chi}{9(\mu_0+\Pi_{I0})}
+\frac{2\Pi_{II0}\chi}{9(\mu_0+\Pi_{II0})}$. The above equation is
known as collapse equation of axisymmetric stellar objects
characterizing meridional and $f(R)$ extra order degrees of freedom.

\section{Instability Regions}

Now we proceed to calculate constraints at which meridional axial
symmetric stellar systems undergo instability window at both N and
pN eras with $f(R)$ background. We also examine the role of
stiffness parameter $\Gamma_1$ in this scenario. We also reduce our
results to previously known limiting cases. The formulation of
instability constraints should be compatible with
Tolman-Oppenheimer-Volkoff (TOV) equation. Such type of equation
constrains the relativistic stellar structure coupled with matter
distribution at the phase of static gravitational equilibrium. In
this respect, Barausse \textit{et al.} \cite{new1} investigated
hydrostatic equilibrium phases of relativistic models by obtaining
modified version of TOV equation in $f(R)$ gravity. Recently,
Astashenok \textit{et al.} \cite{new2} calculated extended version
of TOV equation with equation of state in the realm of cubic as well
as quadratic corrections and found that such an equation can be used
to describe viable models of compact objects. Here, we formulate TOV
equation that will help us to obtain some limits on fluid-energy
density and its derivatives to avoid curvature divergence at the
stellar boundary. The $11$ and $22$ components of metric $f(R)$
field equations, respectively, provide
\begin{align}\label{124}
&\frac{A'}{A}=\frac{B^2}{\gamma}\left[\frac{\kappa}{f_R}\left(
P+\frac{\Pi_{I}}{3}+\frac{\xi_2}{{\kappa}B^2}\right)-\frac{\xi_1}
{\Delta^2B^2}\right],\\\label{125}
&\frac{A_\theta}{A}=\frac{1}{\gamma_1}\left[\frac{\kappa}{f_R}\left\{
\frac{{\mu}L^2}{A^2}+\frac{\Delta}{A^2}\left(P+\frac{2}{9}\left(
\Pi_{II}+2\Pi_I\right)\right)+\frac{\xi_4}{\kappa}\right\}
-\frac{\xi_3}{4A^2}\right],
\end{align}
where
\begin{align}\label{126}
\gamma&=\frac{A^2B^2r^2}{\Delta}\left[\frac{16}{\Delta}\left(1
+\frac{rC'}{C}+\frac{rB'}{B}
\right)+\frac{rf'_R}{f_R}\right],\\\nonumber
\gamma_1&=\frac{A^2B^2r^2L^2}{\Delta^2}\left(\frac{r^2B^2\dot{C}}{LC}
-\frac{2B_\theta}{B}2\frac{2C_\theta}{C}\right)-\frac{r^2B^2A^2f_
{R\theta}}{{\kappa}f_R}\left(\frac{r^2A^2B^2}{\Delta}-\frac{L^2}{\Delta}\right)\\\label{128}
&-\frac{B^4A^4r^4}{\Delta^2}
\left(\frac{C_\theta}{C}+\frac{B_\theta}{B}\right),\\\nonumber
\xi_1&=\Delta^2G_{11}-4{\Delta}r^2A^2b^2\frac{A'}{A}\left(\frac{4}{r}
+\frac{4C'}{C}+\frac{4B'}{B}\right),\quad\xi_2=\overset{(D)}{T_{11}}
+\frac{A^2B^2r^2f'_RA'}{{\kappa}A\Delta},\\\nonumber
\xi_3&=4\Delta^2G_{22}+4B^4A^4r^4\left(\frac{A_{\theta}C_\theta}{AC}+
\frac{A_{\theta}B_\theta}{AB}\right)-4A^2B^2r^2L^2\left(\frac{r^2B^2
\dot{C}A_\theta}{LCA}-\frac{2A_{\theta}B_\theta}{AB}
\right.\\\nonumber
&\left.-\frac{2A_{\theta}C_\theta}{AC}\right),\quad\xi_4=\overset{(D)}
{T_{22}}-\frac{A^2B^2r^2f_{R\theta}}{\Delta}
\left(\frac{r^2B^2AA_\theta}{\Delta}-\frac{L^2A_\theta}{A\Delta}\right).\\\nonumber
\end{align}
The corresponding Misner-Sharp mass function \cite{new10} takes the
form
\begin{align}\label{140}
m_{tot}=\frac{r^3B}{2}\left(\frac{r^2B^2\dot{B}^2}{\Delta}-\frac{B'}{rB^2}-\frac{2B'}{rB}
-\frac{A^2B^2_\theta}{\Delta}-\frac{2LB_\theta\dot{B}}{\Delta}\right).
\end{align}
Using Eq.(\ref{140}), second and third laws of conservation of usual
energy-momentum tensor as well as Eqs.(\ref{124}) and (\ref{125}),
we obtain TOV equations
\begin{align}\nonumber
-P'&=\left[\frac{9{\mu}r^4A^4\psi_m^4-9r^2A^2\psi_m^2\Delta
\{9P+2(\Pi_{II}+2\Pi_I)\}}{9\Delta^2}\right]
\frac{\psi_m^2}{\gamma}\left[\frac{\kappa}{f_R}\left(
P+\frac{\Pi_{I}}{3}\right.\right.\\\nonumber
&\left.\left.+\frac{\xi_2}{{\kappa}\psi_m^2}\right)-\frac{
\xi_1}{\Delta^2\psi_m^2}\right]
+\left[\xi_5+\frac{2}{9}(2\Pi_{II}+\Pi_I)\right],\\\nonumber
-P_\theta&=\left[\frac{-5{\mu}r^2A^4\psi_m^2L-r^2A^2\psi_m^2
\{9P+2(\Pi_{II}+2\Pi_I)\}}{9\Delta^2}\right]
\frac{\Delta}{\gamma_1A^2}\left[\frac{\kappa}{f_R}\left\{
\frac{{\mu}L^2}{A^2}\right.\right.\\\nonumber
&\left.\left.+\frac{\Delta}{A^2}
\left(P+\frac{2}{9}(\Pi_{II}+2\Pi_I)\right)+\frac{\xi_4}{\kappa}\right\}
+\frac{\xi_1}{4A^2}\right]+\frac{\xi_7\Delta}{A^2}.
\end{align}
where
\begin{align}\nonumber
\xi_5&=T^{0\beta}_{~~;\beta}-\left(P+\frac{2}{9}(2\Pi_{II}+\Pi_I)\right)
\frac{r^2A^2B^2A'}{{\Delta}A}+\frac{A'{\mu}r^4A^4B^4}{\Delta^2A}-
\left(P+\frac{2}{9}(\Pi_{II}+2\Pi_I)\right)',\\\nonumber
\xi_6&=\xi_2+\frac{A^2r^2B^2f'_R}{\Delta}\left(\frac{B'}{B}+\frac{1}{r}\right)
-\frac{C'f'_R}{C}-\frac{L\Delta L'f'_R}{2},\\\nonumber
\xi_7&=T^{1\beta}_{~~;\beta}-\frac{A^2}{\Delta}P_{\theta}+\frac{A_\theta}{A}
\left[\frac{5r^2A^4B^2L\mu}{\Delta^2}-\frac{r^2A^2B^2}{\Delta^2}
\left(P+\frac{2}{9}(\Pi_{II}+2\Pi_I)\right)\right],\\\nonumber
\psi_m&=\frac{(2m_{tot}-r^2B')+D\sqrt{(r^2B'-2m_{tot})^2+4r^4\dot{B}^2(rA^2B'
-A^2B^2_\theta-LB_\theta\dot{B})}}{2\dot{B}^2r^3},
\end{align}
where $\psi_m$ is calculated by taking an assumption that reflection
effects are far lesser than that produced by other scale factors in
the evolution of axisymmetric system. In order to examine the
contributions of $f'_R$ and $f_{R\theta}$ across the meridional
non-static axial relativistic object, we multiply both sides of the
above equations with $\frac{df}{dP}$. After some manipulations, this
yields couple of quadratic equations in $f'_R$ and $f_{R\theta}$
whose solutions become
\begin{align}\nonumber
f'_R&=\frac{1}{18f_RCr^3A^2\psi^2_m\Delta}\left[-\psi_mr^2A^2(144(f_R
\psi_mC+f_R\psi_mrC'+Cr\psi'_m)\right.\\\label{129}
&\left.+2\mathcal{C}_1
{\Delta}rC\psi_m+9\mathcal{C}_2rC\psi_m\Delta)+D\sqrt{\Delta_1}\right],\\\nonumber
f_{R\theta}&=\frac{1}{72A^4B^3CL\Delta(A^2b^2r^2-L^2)}\left[-36A^4f_RB^2r^2L^2\kappa
(r^2B^3\dot{C}-2CLb_\theta\right.\\\nonumber
&\left.-2LBC_\theta)+36A^6B^4Lr^4f_R\kappa
(BC_\theta-CB_\theta)-36A^2r^2B^2LC\Delta^2(r^2B^2A^2\right.\\\label{130}
&\left.-L^2)+D\sqrt{\Delta_2}\right],
\end{align}
where $\Delta_1$ and $\Delta_2$ are discriminants of $f'_R$ and
$f_{R\theta}$ quadratic equations,
$\mathcal{C}=\left[\frac{9{\mu}r^4A^4\psi_m^4-9r^2A^2\psi_m^2\Delta
\{9P+2(\Pi_{II}+2\Pi_I)\}}{9\Delta^2}\right]
\frac{df_R}{dP},~\mathcal{C}_1=2(\Pi_{II}+2\Pi_I)'\frac{df_R}{dP}$,
while $D=\pm1$. We shall take $A_0=1-\varphi,~B_0=1+\varphi$ with
$\varphi=\frac{m_0}{r}$ for pN epochs, therefore
\begin{align}\nonumber
\frac{A'_0}{A_0}=(1+\varphi)'(1-\varphi),\quad
\frac{A_{0\theta}}{A_0}=(1+\varphi)_\theta(1-\varphi).
\end{align}
Over the surface of axial reflection relativistic star object,
Eqs.(\ref{129}) and (\ref{130}) yield
\begin{align}\label{131}
\frac{f'_R}{f_R}=\frac{W_1(D-1)}{18C\psi_m\Delta},\quad
\frac{f_{R\theta}}{f_R}=\frac{W_2(D-1)}{2},
\end{align}
where $W_1=144(\psi_mC+\psi_mrC'+rC\psi'_m)$ and
$W_2=\frac{A^2r^2L\kappa}{(A^2B^2r^2-L^2)}(r^2\dot{C}\psi_m+2LC\psi_{m\theta}
-2LC_\theta\psi_m)-A^4B^2r^4\kappa(BC_\theta-CB_\theta)+r^2C\Delta^2(r^2A^2B^2-L^2)$.
It can analyzed from Eq.(\ref{131}) that on setting $D=-1$, one can
get specific forms of $\gamma$ and $\gamma_1$ from Eqs.(\ref{126})
and (\ref{128}) which will make $\frac{A'}{A}$ and
$\frac{A_\theta}{A}$ approach to $\infty$ with
$(r,\theta)\rightarrow(r^-,\theta^-)$, while finite value of
$\frac{A'}{A}$ and $\frac{A_\theta}{A}$ can be achieved for
$(r,\theta)\rightarrow(r^+,\theta^+)$. For physically viable stellar
model, we take $D=1$ which yields $f_{R\theta}=0=f'_R$ for
$(r,\theta)\rightarrow(r^-,\theta^-)$. This reinforces the
continuity of $f_{R\theta},~f'_R$ as well as $A'$ over the surface
of axial stellar structure with reflection degrees of freedom.

\subsection{Newtonian Approximation}

In order to evaluate instability conditions at N regime, we take
$A_0=B_0=1$ and assume anisotropic pressure to be less than zero
which is the criterion for collapsing celestial body. We also take
configurations of initial perturbed structural coefficients to be
$C_0=L_0=r$. Consequently, the collapse equation (\ref{40}) turns
out to be
\begin{align}\nonumber
&\Gamma_1\phi'_{\mathcal{N}}+\frac{9}{4r}\phi_{\mathcal{N}}\Gamma_1-\Gamma_1\frac
{\Pi_{KN0,\theta}}{2r\sqrt{2}}\left(\frac{2c}{r}+3b+\frac{l}{r}
\right)_\theta=\frac{3}{8r}\left(\frac{2c}{r}+3b+\frac{l}{r}\right)
-2b\left[P_0'\right.\\\nonumber
&\left.+\frac{2}{9}(2\Pi'_{I0}+\Pi'_{II0})\right]+\frac{1}{2r\sqrt{2}}
\left(2b-\frac{l}{r}\right)+\frac{b}{2}X_{30{\mathcal{N}}}+\frac{l}
{2r}X_{10{\mathcal{N}}}+\Upsilon+\Phi+\zeta,
\end{align}
where subscript ${\mathcal{N}}$ indicates the evaluation of term at
N regime. We assume that all terms on both sides of the above
equation are positive. The instability constraint for meridional
axisymmetric fluid configurations is given by
\begin{eqnarray}\label{42}
\Gamma_1<\frac{\frac{3}{8r}\left(\frac{2c}{r}+3b+\frac{l}{r}\right)
-2b\left[P_0'+\frac{2}{9}(2\Pi'_{I0}+\Pi'_{II0})\right]+\phi_1+\zeta_N}{\phi'_{\mathcal{N}}
+\frac{9}{4r}\phi_{\mathcal{N}}-\frac
{\Pi_{KN0,\theta}}{2r\sqrt{2}}\left(\frac{2c}{r}+3b+\frac{l}{r}
\right)_\theta},
\end{eqnarray}
where
$\phi_1=\frac{b}{2}X_{30{\mathcal{N}}}+\frac{l}{2r}X_{10{\mathcal{N}}}
+\frac{1}{2r\sqrt{2}}\left(2b-\frac{l}{r}\right)
+\Upsilon_{\mathcal{N}}+\Phi_{\mathcal{N}}+\zeta_{\mathcal{N}}$. The
system would be in complete hydrostatic equilibrium, if (during
evolution) it can take a value equal to the right hand side of the
above expression. However, on satisfying the above inequality, the
system will move in the unstable phase. This constraint is being
mentioned through $\Gamma_1$ parameter thereby emphasizing the
importance of matter stiffness factor in our investigation.

\subsection{Post-Newtonian Approximation}

Here, we take axial structural coefficients for pN eras and consider
our outcomes upto $O(\varphi)$. Using these relations in
Eq.(\ref{40}), one can have modified collapse equation at pN limit.
This leads to instability inequality through stiffness parameter
\begin{eqnarray}\label{43}
\Gamma_1<\frac{r^2(1-4\phi)\{\varphi'+\frac{1}{r}(1-\varphi)
(1-\varphi)_\theta\}\chi_{pN}+(1-2\varphi)\frac{\chi_{pN}}
{4}(\varphi'+\frac{3}{2r})+\zeta_1}{(1-2\varphi)\phi'_{pN}
+(1-2\varphi)\phi_{pN}[\frac{7}{4r}+\frac{1}{2}(\frac{1}{r}-\varphi')]
+\zeta_2},
\end{eqnarray}
where
\begin{align}\nonumber
&\zeta_1=\Pi_{KN0\theta}-2b(1-3\varphi)\left[P'_0+\frac{2}{9}
(2\Pi'_{I0}+\Pi'_{II0})\right]+\frac{(1+2\varphi)}{r\sqrt{2}}
\left(3b-a+2a\varphi\right.\\\nonumber
&\left.-3b\varphi-\frac{3l}{r}\right)+\frac{l}{2r}X_{10pN}
+\frac{b(1-\varphi)}{2}X_{30pN}+\Upsilon_{pN}+\Phi_{pN}+\zeta_{pN},\\\nonumber
&\zeta_2=-\frac{r^3(1-\varphi)}{2\sqrt{2}}\frac{\Pi_{KN0}
\chi_{pN}}{\mu_0+\Pi_{KN0}}\left[6(1+\varphi)_\theta(1
-\varphi)+(1+\varphi)_\theta(1+\varphi)+2(1+\varphi)\right.\\\nonumber
&\left.\times\{(1-\varphi)_\theta(1+\varphi)+(1
+\varphi)_\theta(1-\varphi)\}\right]+\frac{(1+2\varphi)}
{2\sqrt{2}}\left(\frac{\Pi_{KN0}\chi_{pN}}{\mu_0+\Pi_{KN0}}\right)_\theta,
\end{align}
the subscript pN represents effects of quantities at pN era. The
quantity $\Upsilon_{pN}$ describes the reflection effects of
non-static axial celestial body about its symmetry axis at pN
approximations. It is worth mentioning here that these constraints
coincide with \cite{e28} in the limit $L\rightarrow0$ for $n=2$.

\section{Instability of Realistic Star Object}

Perturbations of stars and black holes have been one of the main
topics of relativistic astrophysics for the last few decades. The
description of such stellar objects has recently attracted various
researchers \cite{new3}. The stability analysis of general
relativistic star process is an important but challenging endeavor.
In such study, the spherical symmetric matter configuration is an
exemplary one. Numerous realistic objects like globular clusters,
galactic bulges and dark matter haloes can be considered as being
roughly spherical geometry. For better understanding of cosmic
censorship hypothesis and hoop conjecture, it is necessary to throw
light on non-spherical collapse. The physical interest in studying
non-spherical symmetries is associated with the fact that
post-shocked clouds are left at the verge of gravitational collapse
forming cylinders or plates at scales of galaxy formation and at
scales of stellar formation in galaxy. For instance, cylindrical
distributions are closely related with the problem of fragmentation
of prestellar clouds \cite{new4}.

We take into account a specific configurations of non-static axial
spacetime. The main purpose is to study instability regimes of
axially symmetric realistic objects that are involved in the
emission of gravitational radiation due to meridional degrees of
freedom. For this purpose, we assume coupling of system with
anisotropic fluid distribution whose energy-momentum tensor is
mentioned in Eq.(\ref{4}). Having arrived at this point, the
relevant question is: does an ideal (or non anisotropic) matter
configuration produce gravitational radiations?

To answer such a burning issue, let us recall that in the seminal
paper of Bondi about the emission of gravitational radiation
(section 6 of \cite{new5}), it is mentioned that for relativistic
dust cloud as well as dissipation-less case of an ideal matter
distribution, the relativistic system cannot be anticipated to
radiate (gravitationally). This is due to the reversible feature of
equation of state as emission of radiation is an irreversible
phenomenon. This happens once when absorption is considered (and/or
Sommerfeld type constraints), which prevents inflow of waves. This
implies that an entropy generator parameter/factor must be present
in the discussion of relativistic source. However, such type of
factor is not present in an ideal fluid and in a collisionless dust
cloud. In particular, the irreversibility of gravitational wave
emissions must be taken in equation of state with the help of an
entropy increasing (dissipative) parameter. In this scenario,
Herrera \textit{et al.} \cite{new6} described a close relationship
between vorticity and gravitational radiations.

We consider the evolution of non-static axisymmetric
self-gravitating system in $f(R)$ gravity and assume that it is in
hydrostatic equilibrium at large past time. Now we want to analyze
that when the phase of equilibrium is disturbed, what happens? Will
this perturbation be relaxed (stable state) or will it grow
(unstable state). In this respect, one needs to take into account
couple of following instabilities
\begin{enumerate}
\item dynamical stability: what happens, if stellar hydrostatic phase is
perturbed?
\item secular (thermal) stability: what happens, when the
state of thermal equilibrium is perturbed?
\end{enumerate}
Since our system is coupled with anisotropic matter configurations
without heat flux, therefore we shall not discuss the second case
and confine ourselves over dynamical instability of relativistic
origin. It is seen that under hydrostatic phase, the stability
criterion is achieved by making linearized field as well as
conservation equations against radial perturbation (14)-(17). It is
remarked that during evolution, the realistic object moves via
several evolutionary patterns determined by instability/stability
degrees of freedom. This suggests that the relativistic systems can
be stable at one instance but not at the other. Thus one needs to
cope with the dynamical evolution of self-gravitating systems by
calculating instability regions at N as well as pN regimes. Such
epochs have vital role in the discussion of gravitational collapse
of compact objects.

The phenomenon of celestial collapse occurs when the state of
hydrostatic equilibrium of a stellar object is disturbed. In
celestial body, nuclear fission reactions occur that start from
hydrogen atoms and produce further complex elements until nuclear
reactions chain stops with iron. These reactions increase the
pressure exerted by gas particles which counterbalance the
gravitational attraction and prevents the star from collapsing.
However, with the passage of time, nuclear reactions decrease as
fuel burns out. Consequently, the necessary pressure becomes
insufficient for a collapsing body to be stable. At this point, the
gravitational force begins to pull matter towards the center of a
body and thus collapse initiates. A celestial body that has
exhausted all its nuclear fuel, can give birth to three possible
compact objects (white dwarfs, neutron stars and black holes) on the
basis of the initial mass of the collapsing body.

It is well-known that, in the scenario of Newtonian regime, the
instability of spherical self-gravitating systems depends purely on
the mean value of stiffness parameter, $\Gamma_1$ \cite{new7} which
is the ratio of fractional Lagrangian variations between pressure
and energy density experienced by matter configurations following
the motion. However, in GR, the stability relies not only on the
average value of $\Gamma_1$ but also on the star radius. However, in
the study of non-static axial reflection system in modified gravity,
the situation is quite different. (It is worthy to stress that we
have assumed $\Gamma_1$ as a constant entity throughout the
analysis). The most important consequence of our study is that,
apart from affirming GR results, $\Gamma_1$ controls emission of
gravitational radiations along with $f(R)$ extra degrees of freedom.
The emission of gravitational radiations causes the loss of both
energy as well as angular momentum which increases the instability
of the meridional axisymmetric object.

More specifically, following the results of Chandrasekhar
\cite{e19}, we deduce that if the anisotropic matter distribution
attains stiffness equal to the right hand side of expressions
(\ref{42}) and (\ref{43}), the system enters into the window of
hydrostatic equilibrium at N and pN regimes. Further, if the
stiffness of fluid increases in such a way that the fractional value
given at the right hand side of (\ref{42}) and (\ref{43}) becomes a
smaller one, then system enters into the stable configurations at
both N and pN approximations, respectively. Dosopoulou \textit{et
al.} \cite{new8} explored the contribution of magnetic fields in the
emergence and existence of vorticity. This strongly suggests that
invoking of magnetic fields in the study of stability of
gravitationally radiating sources deserves attention for future
work.

\section{Conclusions}

It is well-known that the most general non-static axial geometry
incorporates reflection (meridional) and rotation effects coming out
from non-diagonal $dtd\theta$ and $dtd\phi$ metric coefficients. In
order to dealt analytically with instability constraints of axially
symmetric spacetime, several attempts have been made by taking
restricted class of axial geometry. In this paper, we have studied
stability analysis of meridional axial stellar structure with $f(R)$
background. We are observing investigation in a metric $f(R)$
gravity which give rise to non-linear fourth order field equations.
We have formulated the collapse equation by using perturbation
scheme in the generalized Euler equation. We assume complete
hydrostatic equilibrium of axial stellar structure at large past
time, i.e., $T(-\infty)=0$.

We have developed instability constraints at N and pN epochs through
stiffness parameter, $\Gamma_1$ using collapse equation. It is found
that axial stellar structure would be unstable until it obeys
relation (\ref{42}) at N regime while relation (\ref{43}) at pN era.
Breaching of these inequalities will eventually move the system
towards stable window. These constraints depend upon adiabatic
index, static combinations of anisotropy, energy density and dark
source corrections due to $R+\epsilon R^n$ model. It is seen that
dark source corrections tend to stabilize structure formation
phenomenon due to its non-attractive behavior while the presence of
non-diagonal terms in instability ranges indicate occurrence of
gravitational radiations which correspond to flow of super-energy
\cite{e18}.

We have found non-vanishing component of vorticity tensor which
corresponds to non-static meridional axial structure coefficient.
The inclusion of non-diagonal scale factor in the stability analysis
leads to interesting phenomenon of gravitational radiations for
$\epsilon R^n$ corrections. These extra-order $f(R)$ corrections
affect the passive gravitational mass which in turn affect the rate
of stellar collapse. We have developed instability constraints
(\ref{42}) and (\ref{43}) with weak field and pN approximations,
respectively. These constraints can be applied to axisymmetric
self-gravitating system with reflection degrees around symmetry axis
at some particular cosmic epochs depending upon the chosen values of
$n$. We can categorize different eras of cosmic dynamics associated
with $\epsilon R^n$ as follows.
\begin{itemize}
\item For $n=2$, the instability constraints for specific model of the type $R+\epsilon R^2$
can be obtained. The existence of $R^2$ correction in the field
equations can be helpful to explain inflationary mechanism of
cosmos. The term $\alpha R^2$ represents accelerated expansion of
the universe. This model is compatible with temperature anisotropies
noticed in cosmic microwave background radiations \cite{e31} and
hence viable for inflationary scalar field models.
\item The choice $n=3$ favors to host significant
massive compact objects coming out from cubic $f(R)$ higher
curvature terms \cite{e25}. This provides realistic signature of the
presence of more massive and huge self-gravitating stellar systems
which have direct correspondence with the observational cosmology.
\item This gravitational dynamics at late-time universe era
can be obtained by substituting $n=-1$ in instability constraints at
both N and pN regimes. It is noticed that gravitational contribution
due to negative curvature power serves as dark energy thereby
supporting current accelerating cosmic epochs \cite{e34}.
\item For $\epsilon=0$, instability constraints for
Einstein gravity can be obtained at both N and pN eras which
describes relatively less stable axial stellar structure.
\end{itemize}

Finally, we remark that supermassive stellar systems survive more
abundantly in extended gravity than in GR as such theories (for
instance $f(R)$ gravity) are more likely to host huge stars with
smaller radii. This leads to the existence of more dense
relativistic systems which have direct relevance with observational
gravitational physics. It is interesting to mention here that all
our results reduce to restricted class of instability analysis
\cite{e28} by neglecting non-diagonal terms and assuming $n=2$ in
$f(R)$ model.

\vspace{0.5cm}

{\bf Acknowledgement}

\vspace{0.25cm}

We would like to thank the Higher Education Commission, Islamabad,
Pakistan for its financial support through the {\it Indigenous Ph.D.
Fellowship for 5K Scholars, Phase-II, Batch-I}.

\vspace{0.25cm}

\renewcommand{\theequation}{A\arabic{equation}}
\setcounter{equation}{0}
\section*{Appendix A}

The extra $f(R)$ curvature terms for Eqs.(\ref{19})-(\ref{21}) are
\begin{align}\nonumber
D_0&=\dot{V}_1+X'_1+X'_3+\left(\frac{3B^2r^2AA'}{\Delta}
+\frac{2A^2B^2r}{\Delta}+\frac{2r^2A^2BB'}{\Delta}
+\frac{C'}{C}+\frac{B'}{B}\right)X_1\\\nonumber &+\left(\frac{
3r^2B^2AA_\theta}{\Delta}+\frac{B_\theta}{B}+\frac{C_\theta}
{C}+\frac{A^2r^2BB_\theta}{\Delta}\right)X_3+\left(\frac{r^2
B^3\dot{B}}{\Delta}+\frac{B'}{B}\right)V_2\\\nonumber
&+\frac{r^2B^2}{\Delta}(r^2B\dot{B}V_3+C\dot{C}V_4)
+\dot{W}_1+(Y_1+Y_3)'+\frac{LW_1}{\Delta}(AA_\theta+L\dot{L})
+\frac{L\dot{L}V_1}{\Delta}\\\nonumber
&+\left(\frac{3r^2B^2AA'}{\Delta}+\frac{B'}{B}+\frac{2rA^2B^2}
{\Delta}+\frac{2r^2A^2BB'}{\Delta}+\frac{C'}{C}+\frac{4LL'}{
\Delta}\right)Y_1\left(\frac{B_\theta}{B}\right.\\\nonumber
&\times\left.\frac{3r^2B^2AA_\theta}{\Delta}+\frac{r^2A^2BB_\theta}{\Delta}+
\frac{C_\theta}{C}+\frac{3Lr^2B\dot{B}}{\Delta}-\frac{r^2B^2L'}
{2\Delta}+\frac{rLB^2}{\Delta}+\frac{r^2LBB'}{\Delta}\right.\\\nonumber
&\left.-\frac{r^2
LB\dot{B}}{\Delta}+\frac{LL_\theta}{\Delta}\right)Y_3+\left(\frac{3r^2LB
\dot{B}}{\Delta}-\frac{r^2B^2L'}{2\Delta}+\frac{r^2LBB'}{\Delta}
+\frac{rB^2L}{\Delta} +\frac{LL_\theta}{\Delta}\right.\\\nonumber
&\left.-\frac{r^2LB
\dot{B}}{\Delta}\right)X_3+\left(\frac{r^2B^3\dot{B}}{\Delta}+\frac{B'}{B}-\frac{LB
B_\theta}{\Delta}\right)W_2-\frac{LBB_\theta}{\Delta}V_2
+\left(\frac{rLB^2}{\Delta}\right.\\\nonumber
&\left.+\frac{r^2BLB'}{\Delta}-\frac{r^2
B^2L'}{2\Delta}\right)(X_2+Y_2)+\left(\frac{rBLB_\theta}{\Delta}-\frac{r^2B^2L_\theta}
{\Delta}+\frac{r^4B^3\dot{B}}{\Delta}\right)W_3\\\label{d0}
&+\left(\frac{rBLB_\theta}{\Delta}-\frac{r^2B^2L_\theta}{\Delta}\right)V_3
+\left(\frac{r^2}{\Delta}B^2C\dot{C}-\frac{LCC_\theta}{\Delta}\right)W_4
-\frac{LCC_\theta}{\Delta}V_4,\\\nonumber
D_1&=\dot{X}_1-\left(r+\frac{r^2B'}{B}\right)W_3-\frac{CC'}{B^2}
W_4+X_{2\theta}+V'_2+\frac{AA'}{B^2}V_1+\left(\frac{2B'}{B}
\right.\\\nonumber
&\left.+\frac{r^2B^2AA'}{\Delta}+\frac{2rA^2B^2}{\Delta}+\frac{2r^2A^2BB'}{\Delta}
+\frac{C'}{C}\right)V_2+\left(\frac{3B_\theta}{B}+\frac{
r^2B^2AA_\theta}{\Delta}\right.\\\nonumber
&\left.+\frac{r^2A^2BB_\theta}
{\Delta}+\frac{C_\theta}{C}\right)X_2+\frac{A^2r^2B\dot{B}}{\Delta}X_3-\left(r+\frac{r^2B'}{B}
\right)V_3-\frac{CC'V_4}{B^2}+\dot{Y}_1\\\nonumber
&+W'_2+Y_{2\theta}
+\left(\frac{3\dot{B}}{B}+\frac{r^2B^2A\dot{A}}{\Delta}
+\frac{\dot{C}}{C}+\frac{LAA_\theta}{\Delta}+\frac{
L\dot{L}}{\Delta}\right)Y_1+\left(\frac{LAA_\theta}{\Delta}\right.\\\nonumber
&\left. +\frac{L\dot{L}}{\Delta}\right)X_1+\left(\frac{2B'}{B}
+\frac{r^2B^2AA'}{\Delta}+\frac{2rA^2B^2}{\Delta}+\frac{2r^2A^2BB'}{\Delta}+\frac{C'}{C}+\frac{3LL'}
{\Delta}\right)W_2\\\nonumber
&+\frac{3LL'}{\Delta}V_2+\left(\frac{3B_\theta}{B}
+\frac{r^2B^2AA_\theta}{\Delta}+\frac{r^2B^2BB_\theta}{\Delta}
+\frac{C_\theta}{C}+\frac{LL_\theta}{\Delta}\right)Y_2
+\frac{LL_\theta}{\Delta}X_2\\\label{d1}
&+\left(\frac{r^2A^2B\dot{B}}{
\Delta}-\frac{L'}{B}-\frac{LAA_\theta}{\Delta}\right)Y_3
-\left(\frac{L'}{B}+\frac{LAA_\theta}{\Delta}\right)X_3,\\\nonumber
D_2&=\dot{X}_3+X'_2+V_{3\theta}+\frac{A^3A_\theta}{\Delta}
V_1-\frac{A^2BB_\theta}{\Delta}V_2-\frac{A^2CC_\theta}{
\Delta}V_4+\left(\frac{r^2B^2AA_\theta}{\Delta}\right.\\\nonumber
&\left.+\frac{2r^2A^2BB_\theta}{\Delta}+\frac{B_\theta}{B}+\frac{C_\theta}{C}\right)V_3
+\left(\frac{3A^2r^2B\dot{B}}{\Delta}-\frac{2ALA_\theta}{
\Delta}+\frac{r^2B^2A\dot{A}}{\Delta}\right.\\\nonumber
&\left.+\frac{\dot{B}}{B}
+\frac{\dot{C}}{C}\right)X_3+Y'_2+\dot{Y}_3
+\left(\frac{6rA^2B^2}{\Delta}+\frac{6r^2A^2BB'}{\Delta}
+\frac{2r^2B^2AA'}{\Delta}+\frac{B'}{B}\right.\\\nonumber
&\left.+\frac{C'}{C}\right)
X_2+W_{3\theta}+\left(\frac{A^3A_\theta}{\Delta}+\frac{A^2
\dot{L}}{\Delta}-\frac{AL\dot{A}}{\Delta}\right)W_1+\left(\frac{A^2
\dot{L}}{\Delta}-\frac{AL\dot{A}}{\Delta}\right)V_1\\\nonumber
&-\left(
\frac{A^2BB_\theta}{\Delta}-\frac{BL\dot{B}}{\Delta}\right)W_2
-\frac{BL\dot{B}}{\Delta}V_2-\left(\frac{A^2}{\Delta}CC_\theta
+\frac{CL\dot{C}}{\Delta}\right)W_4\\\nonumber
&-\frac{CL\dot{C}}{\Delta}V_4+\left(\frac{r^2B^2AA_\theta}{\Delta}
+\frac{2r^2A^2BB_\theta}{\Delta}+\frac{r^2BL\dot{B}}{\Delta}
+\frac{2LL_\theta}{\Delta}+\frac{B_\theta}{B}+\frac{C_\theta}{C}\right.\\\nonumber
&\left.-\frac{2r^2LB
\dot{B}}{\Delta}\right)W_3+\frac{LV_3}{\Delta}(r^2BL\dot{B}
+2L_\theta-2r^2B\dot{B})+\left(\frac{A^2L'}{\Delta}-\frac{
2ALA'}{\Delta}\right)\\\nonumber
&\times(X_1+Y_1)+\left(\frac{3A^2r^2B\dot{B}}{\Delta}-\frac{
2ALA_\theta}{\Delta}+\frac{r^2B^2A\dot{A}}{\Delta}+\frac{\dot{
B}}{B}+\frac{\dot{C}}{C}+\frac{L\dot{L}}{\Delta}\right)Y_3
\\\nonumber
&+\left(\frac{L\dot{L}}{\Delta}-\frac{2LAA_\theta}{\Delta}\right)X_3+\left(\frac{6
rA^2B^2}{\Delta}+\frac{6r^2A^2BB'}{\Delta}+\frac{2r^2B^2AA'}
{\Delta}+\frac{B'}{B}\right.\\\label{d2}
&\left.+\frac{C'}{C}+\frac{7LL'}{2\Delta}\right)Y_2+\frac{7LL'}{2\Delta}X_2.
\end{align}

The perturbed parts of Eq.(\ref{19}) are
\begin{align}\nonumber
g&=\left[x'_1+x_1\left(\frac{3B_0^2r^2A_0A'_0}{\Delta_0}+\frac{B'_0}{B_0}
+\frac{2A_0^2B_0^2r}{\Delta_0}+\frac{2A_0^2r^2B_0B'_0}{\Delta_0}
+\frac{C'_0}{C_0}\right)+X_{10}\right.\\\nonumber
&\times\left\{\frac{3B_0^2r^2A_0A'_0}{\Delta_0}\left(\frac{2b}{B_0}
+\frac{a}{A_0}+\frac{a'}{A'_0}-\frac{d}{\Delta_0}\right)+\left(\frac
{b}{B_0}\right)'+\frac{2A_0^2B_0^2r}{\Delta_0}\left(\frac{2a}{A_0}
+\frac{2b}{B_0}\right.\right.\\\nonumber
&\left.\left.-\frac{d}{\Delta_0}\right)+\frac{2A_0^2r^2B_0B'_0}{
\Delta_0}\left(\frac{2a}{A_0}+\frac{b}{B_0}+\frac{b'}{B'_0}\right)
+\left(\frac{c}{C_0}\right)'\right\}+x_3\left(\frac{3r^2B_0^2A_0
A_{0\theta}}{\Delta_0}\right.\\\nonumber
&\left.+\frac{C_{0\theta}}{C_0}+\frac{B_{0\theta}}{B_0}+\frac{A_0^2
r^2B_0^2B_{0\theta}}{\Delta_0}\right)+X_{30}\left\{\frac{3r^2B_0^2
A_0A_{0\theta}}{\Delta_0}\left(\frac{2b}{B_0}+\frac{a}{A_0}+\frac{
a_\theta}{A_{0\theta}}\right.\right.\\\nonumber
&\left.\left.-\frac{d}{\Delta_0}\right)+\left(\frac{c}{C_0}\right)_\theta+\left(\frac{c}{C_0}
\right)_\theta+\frac{A_0^2r^2B_0^2B_{0\theta}}{\Delta_0}\left(
\frac{2a}{A_0}+\frac{2b}{B_0}+\frac{b_\theta}{B_{0\theta}}\right)
\right\}\\\nonumber &+W_{10}\frac{L_0A_0A_{
0\theta}}{\Delta_0}\left(\frac{l}{L_0}+\frac{a}{A_0}+\frac{
a_\theta}{A_{0\theta}}-\frac{d}{\Delta_0}\right)+y_1\left(
\frac{3r^2B_0^2A_0^2B_{0\theta}}{\Delta_0}+\frac{C'_0}{C_0}+\frac{B'_0}{B_0}\right.\\\nonumber
&\left.+\frac{2A_0^2B_0^2r}{\Delta_0}+\frac{
2A_0^2B_0r^2B'_0}{\Delta_0}+\frac{4L_0L'_0}{
\Delta_0}\right)+Y_{10}\left\{\frac{3r^2B_0^2A_0^2B_{0\theta}}
{\Delta_0}\left(\frac{2b}{B_0}+\frac{a}{A_0}+\frac{a'}{A'_0}\right.\right.\\\nonumber
&\left.\left.-\frac{d}{\Delta_0}
\right)+\left(\frac{b}{B_0}\right)'+\frac{2A_0^2B_0^2r}{\Delta_0}
\left(\frac{2a}{A_0}+\frac{2b}{B_0}-\frac{d}{\Delta_0}\right)
+\frac{2A_0^2B_0r^2B'_0}{\Delta_0}
\left(\frac{2a}{A_0}+\frac{b}{B_0}\right.\right.\\\nonumber
&\left.\left.+\frac{b'}
{B'_0}-\frac{d}{\Delta_0}\right)+\left(\frac{c}{C_0}\right)'
+\frac{4L_0L'_0}{\Delta_0}\left(\frac{l}{L_0}+\frac{l'}{L'_0}
-\frac{d}{\Delta_0}\right)\right\}+\frac{4x_1L_0L'_0}{\Delta_0}
+X_{10}\\\nonumber
&\times\frac{4L_0L'_0}{\Delta_0}\left(\frac{l}{L_0}+\frac{l'}
{L_0}-\frac{d}{\Delta_0}\right)+y_3\left(\frac{3r^2B_0^2A_0
A_{0\theta}}{\Delta_0}+\frac{C_{0
\theta}}{C_0}+\frac{A_0^2r^2B_0B_{0\theta}}{\Delta_0}\right.\\\nonumber
&\left.+\frac{B_{0\theta}}{B_0}
-\frac{r^2L'_0B_0^2}{2\Delta_0}+\frac{rL_0B_0^2}{
\Delta_0}+\frac{r^2L_0B_0B'_0}{\Delta_0}+\frac{L_0L_{0
\theta}}{\Delta_0}\right)+Y_{30}\left\{\frac{3r^2B_0^2A_0
A_{0\theta}}{\Delta_0}\right.\\\nonumber
&\left.\times\left(\frac{2b}{B_0}+\frac{a}{A_0}+\frac{a_\theta}{A_{0\theta}}-\frac{d}{
\Delta_0}\right)+\left(\frac{b}{B_0}\right)_\theta+\frac{
A_0^2r^2B_0B_{0\theta}}{\Delta_0}\left(\frac{b}{B_0}
+\frac{2a}{A_0}+\frac{b_\theta}{B_{0\theta}}\right.\right.\\\nonumber
&\left.\left.-\frac{d}
{\Delta_0}\right)+\left(\frac{c}{C_0}\right)_\theta
-\frac{r^2L'_0B_0^2}{2\Delta_0}\left(\frac{l'}
{L'_0}+\frac{2b}{B_0}-\frac{d}{\Delta_0}\right)+\frac{
rL_0B_0^2}{\Delta_0}\left(\frac{l}{L_0}+\frac{2b}{B_0}
-\frac{d}{\Delta_0}\right)\right.\\\nonumber
&\left.+\frac{r^2L_0B_0B'_0}{\Delta_0}
\left(\frac{l}{L_0}+\frac{b}{B_0}+\frac{b'}{B'_0}-\frac{d}
{\Delta_0}\right)+\frac{L_0L_{0\theta}}{\Delta_0}
\left(\frac{l}{L_0}+\frac{l_\theta}{L_{0\theta}}-\frac{d}
{\Delta_0}\right)\right\}+x_3\\\nonumber
&\times\left(\frac{rB_0^2L_0}
{\Delta_0}-\frac{r^2B_0^2L'_0}{2\Delta_0}+\frac{r^2B_0B'_0L_0}{\Delta_0}+\frac{L_0L_{
0\theta}}{\Delta_0}\right)+X_{30}\left\{\frac{rB_0^2
L_0}{\Delta_0}\left(\frac{2b}{B_0}-\frac{d}{\Delta_0}
\right)\right.\\\nonumber
&\left.-\frac{r^2B_0^2L'_0}{2\Delta_0}\left(\frac{2b}
{B_0}+\frac{l'}{L'_0}-\frac{d}{\Delta_0}\right)+\frac{
r^2B_0B'_0L_0}{\Delta_0}\left(\frac{b}{B_0}+\frac{l}{L_0}
+\frac{b'}{B'_0}-\frac{d}{\Delta_0}\right)\right.\\\nonumber
&\left.+\frac{L_0 L_{0\theta}}{\Delta_0}\left(\frac{l}{L_0}+\frac{
l_\theta}{L_{0\theta}}-\frac{d}{\Delta_0}\right)\right\}
+w_2\left(\frac{B'_0}{B_0}-\frac{L_0B_0B_{0\theta}}{
\Delta_0}\right)+\left\{\left(\frac{b}{B_0}\right)'
-\frac{L_0}{\Delta_0}\right.\\\nonumber
&\left.{\times}B_0B_{0\theta}\left(\frac{l}{L_0}
+\frac{b}{B_0}+\frac{b_\theta}{B_{0\theta}}-\frac{d}
{\Delta_0}\right)\right\}W_{20}-V_{20}\frac{L_0B_0B_{0\theta}}{\Delta_0}
\left(\frac{l}{L_0}+\frac{b}{B_0}+\frac{b_\theta}{B_{0
\theta}}\right.\\\nonumber
&\left.-\frac{d}{\Delta_0}\right)-v_2\frac{L_0B_0B_{0
\theta}}{\Delta_0}+(x_2+y_2)\left(\frac{r}{\Delta_0}
L_0B_0^2+\frac{r^2B_0B'_0}{\Delta_0}-\frac{r^2
B_0^2L'_0}{2A_0}\right)\\\nonumber
&+(x_2+y_2)\left\{\frac{rL_0B_0^2}
{\Delta_0}\left(\frac{l}{L_0}+\frac{2b}{B_0}-\frac{d}{
\Delta_0}\right)+\frac{r^2B_0B'_0}{\Delta_0}
\left(\frac{l}{L_0}+\frac{b}{B_0}+\frac{
b'}{B'_0}\right.\right.\\\nonumber
&\left.\left.-\frac{d}{\Delta_0}\right)-\frac{r^2B_0^2L'_0}
{2A_0}\left(\frac{l'}{L'_0}+\frac{2b}{B_0}-\frac{d}{
\Delta_0}\right)\right\}+(w_3+v_3)\left(\frac{r}{\Delta_0}
B_0L_0B_{0\theta}\right.\\\nonumber
&\left.-\frac{r^2B_0^2L_{0\theta}}
{\Delta_0}\right)+(W_{30}+V_{30})\left\{\frac{rB_0L_0
B_{0\theta}}{\Delta_0}\left(\frac{l}{L_0}+\frac{b}{B_0}+\frac{
b_\theta}{B_{0\theta}}-\frac{d}{\Delta_0}\right) -\left(\frac{
2b}{B_0}\right.\right.\\\nonumber
&\left.\left.+\frac{a_\theta}{A_{0\theta}}-\frac{d}{\Delta_0}
\right)\frac{r^2B_0^2L_{0\theta}}{\Delta_0}\right\}-(v_4+w_4)\frac{L_0C_0C_{0\theta}}{\Delta_0}
-(V_{40}+W_{40})\frac{L_0C_0C_{0\theta}}{\Delta_0}
\\\label{gg}
&\left.\left(\frac{l}{L_0}+\frac{c}{C_0}+\frac{c_\theta}
{C_{0\theta}}-\frac{d}{\Delta_0}\right)+V_{20}
\left(\frac{b}{B_0}\right)'+v_2\frac{B'_0}{B_0}
+\frac{w_1L_0A_0A_{0\theta}}{\Delta_0}\right]T,\\\nonumber
h&=v_1+V_{10}\left(\frac{B_0^2r^2A_0a}{\Delta_0}+\frac{c}
{C_0}\right)+\frac{lL_0}{\Delta_0}(V_{10}+W_{10})+\frac{bB_0^3r^2}{\Delta_0}
(V_{20}+W_{20})\\\label{hh}
&+\frac{bB_0^3r^4}{\Delta_0}(V_{30}+W_{30})+\frac{cC_0B_0^2r^2}{\Delta_0}(V_{40}+W_{40})
+w_1+\frac{L_0^2lW_{10}}{\Delta_0}.
\end{align}

The perturbed parts of Eq.(\ref{38}) are
\begin{align}\nonumber
\zeta&=\frac{r^3A_0B_0^3}{\Delta_0^{\frac{3}{2}}}\left(\frac{
a}{A_0}+\frac{3b}{B_0}-\frac{3d}{\Delta_0}\right)\left\{\frac{
A_{0\theta}}{A_0}+\frac{6B_{0\theta}}{B_0}+\frac{C_{0\theta}}
{C_0}+\frac{4r^2A^2_0B^2_0}{\Delta_0}\left(\frac{A_{0\theta}
}{A_0}+\frac{B_{0\theta}}{B_0}\right)\right\}.\\\nonumber
&+\frac{r^3A_0B_0^3}{\Delta^{3/2}_0}{\Pi}_{KL0}\left[\frac{6
B_{0\theta}}{B_0}\left(\frac{b_\theta}{B_{0\theta}}+\frac{b}
{B_0}\right)+\left(\frac{a}{A_0}\right)_\theta+\left(\frac{c}
{C_0}\right)_\theta+\frac{4r^2A_0^2B_0^2}{\Delta_0^2}\left(\frac{
2a}{A_0}\right.\right.\\\nonumber
&\left.\left.+\frac{2b}{B_0}-\frac{2d}{\Delta_0}\right)\left(
\frac{a}{A_0}+\frac{b}{B_0}\right)_\theta\right]-\frac{\mu_0
r^4A_0^4}{\Delta_0^2}\left(\frac{a}{A_0}\right)'-\left\{{P_0}
+\frac{2}{9}(2{\Pi}_{I0}+{\Pi}_{II0})\right\}\frac{1}{B_0^2}\\\nonumber
&\times\left[\left(\frac{c}{C_0}\right)'+\frac{r^2A_0^2B_0^2}
{Z_0^2}\left(\frac{2a}{A_0}+\frac{2b}{B_0}-\frac{2d}{\Delta_0}\right)
\left(\frac{A_0'}{A_0}+\frac{1}{r}\right)+\frac{r^2A_0^2B_0^2}
{\Delta_0}\left(\frac{a}{A_0}+\frac{b}{B_0}\right)'\right]
\\\label{zeta}
&-\frac{2b}{B_0^2}\left\{{P_0}+\frac{2}{9}(2{\Pi}_{I0}+{
\Pi}_{II0})\right\}\left[\frac{C_0'}{C_0}+\frac{r^2A_0^2
B_0^2}{\Delta_0}\left(\frac{A_0'}{A_0}+\frac{1}{r} \right)
\right],\\\nonumber
\Upsilon&=-\left[\frac{4r^3A_0L_0B_0^3L'_{0\theta}}{\Delta^{5/2}}
\left(\frac{a}{A_0}+\frac{3b}{B_0}-\frac{3d}{\Delta_0}\right)+
\frac{4L_0L_{0\theta}}{\Delta}\left(\frac{l}{L_0}+\frac{l_\theta}
{L_{0\theta}}-\frac{d}{\Delta_0}\right)-\frac{A_{0\theta}L_0}
{r^2A_0B_0^2}\right.\\\nonumber
&\times\left.\left(\frac{l}{L_0}+\frac{a_\theta}{A_{0\theta}}
-\frac{a}{A_0}-\frac{2b}{B_0}\right)+\frac{\mu_0L_0^2A_0^2r^2}{\delta_0^2}
\left(\frac{2l}{L_0}+\frac{2a}{A_0}-\frac{2d}{\Delta_0}\right)
+\frac{\mu_0L_0^2A_0^2r^2}{\Delta_0^2}
\left\{\frac{L'_0}{2L_0}\right.\right.\\\nonumber
&\left.\left.\times\left(\frac{l'}{L'_0}-\frac{l}{L_0}\right)
+\left(\frac{b}{B_0}\right)'\right\}-\frac{3L_0L'_0}{2B_0^2
\Delta_0}\left\{{P_0}+\frac{2}{9}(2{\Pi}_{I0}+{\Pi}_{II0})\right\}
\left(\frac{l}{L_0}+\frac{l'}{L'_0}-\frac{d}{\Delta_0}\right)\right.\\\nonumber
&\left.+Y_{10}\left(\frac{3b}{B_0}+\frac{ar^2A_0B_0^2}{\Delta_0}
+\frac{c}{C_0}\right)+y_1\right]+\left[\frac{r^2B_0L_0b}{\Delta_0}
+\frac{L_0L_{0\theta}}{\Delta_0}\left(\frac{l}{L_0}+\frac{
l_\theta}{L_{0\theta}}-\frac{d}{\Delta_0}\right)\right.\\\nonumber
&\left.-\frac{r^2L_0bB_0}{\Delta_0}\right]{\omega}X_{20}
-\frac{{\omega}r^2A_0^2bB_0Y_{30}}{\Delta_0}+\frac{A_0L_0
A_{0\theta}}{\Delta_0}\left(\frac{a_\theta}{A_{0\theta}}
+\frac{a}{A_0}+\frac{l}{L_0}-\frac{d}{\Delta_0}\right)Y_{10}
+\frac{L_0}{\Delta_0}\\\nonumber
&{\times}A_0A_{0\theta}(y_1+x_1)+X_{10}\frac{L_0A_0A_{0
\theta}}{\Delta_0}\left(\frac{a_\theta}{A_{0\theta}}+\frac{
a}{A_0}+\frac{l}{L_0}-\frac{d}{\Delta_0}\right)+w_2\left(
\frac{2B'_0}{B_0}+\frac{B_0^2r^2A_0}{\Delta_0}\right.\\\nonumber
&\left.{\times}A'_0+\frac{2rA_0^2B_0^2}{\Delta_0}+\frac{2A_0^2
r^2B_0B'_0}{\Delta_0}+\frac{C'_0}{C_0}+\frac{3L_0L'_0}{\Delta_0}
\right)+W_{20}\left\{2\left(\frac{b}{B_0}\right)'+\frac{B_0^2
r^2A_0A'_0}{\Delta_0}\right.\\\nonumber
&\left.\times\left(\frac{a}{A_0}+\frac{a'}{A'_0}+\frac{2b}{B_0}
-\frac{d}{\Delta_0}\right)+\frac{2rA_0^2B_0^2}{\Delta_0}
\left(\frac{2b}{B_0}+\frac{2a}{A_0}-\frac{d}{\Delta_0}\right)
+\frac{2r^2A_0^2B_0B'_0}{\Delta_0}\left(\frac{2a}{A_0}\right.\right.\\\nonumber
&\left.\left.+\frac{b'}{B'_0}
+\frac{b}{B_0}-\frac{d}{\Delta_0}\right)+\left(\frac{c}{C_0}\right)'
+\frac{6L_0L'_0}{\Delta_0}\left(\frac{l}{L_0}-\frac{d'}{\Delta_0}
+\frac{l'}{L'_0}\right)\right\}+\frac{3w_2L_0L'_0}{\Delta_0}+y_2\\\nonumber
&\times\left(\frac{3B_{0\theta}}{B_0}+\frac{C_{0\theta}}{C_0}
+\frac{r^2B_0^2A_0A_{0\theta}}{\Delta_0}+\frac{r^2A_0^2B_0B_{0
\theta}}{\Delta_0}+\frac{L_0L_{0\theta}}{\Delta_0}\right)+y_{2\theta}+Y_{20}
\left\{3\left(\frac{b}{B_0}\right)_\theta\right.\\\nonumber
&\left.+\frac{r^2B_0^2A_0A_{0\theta}}{\Delta_0}\left(\frac{a}
{A_0}+\frac{a_\theta}{A_{0\theta}}+\frac{2b}{B_0}-\frac{d}{
\Delta_0}\right)+\frac{A_0^2r^2B_0B_{0\theta}}{\Delta_0}
\left(\frac{2a}{A_0}+\frac{b}{B_0}+\frac{b_\theta}{B_{0\theta}}
-\frac{d}{\Delta_0}\right)\right.\\\nonumber
&\left.+\left(\frac{c}{C_0}
\right)_\theta+\frac{L_0L_{0\theta}}{\Delta_0}\left(\frac{l}{L_0}
+\frac{l_\theta}{L_{0\theta}}-\frac{d}{\Delta_0}\right)\right\}
+x_2\frac{L_0L_{0\theta}}{\Delta_0}-y_3\left(\frac{L'_0}{B_0^2}
+\frac{L_0A_0A_{0\theta}}{\Delta_0}\right)\\\nonumber
&+w'_2-Y_{30}\left\{\frac{L'_0}{B_0^2}+\frac{L_0A_0A_{0\theta}}{\Delta_0}
\left(\frac{a}{A_0}+\frac{a_\theta}{A_{0\theta}}+\frac{l}{L_0}
-\frac{d}{\Delta_0}\right)\right\}-r^2W_{30}
\left(\frac{b}{B_0}\right)'-x_3\\\nonumber
&\times\left(\frac{L'_0}{B_0^2}+\frac{L_0A_0A_{0\theta}}{\Delta_0}
\right)-X_{30}\left\{\frac{L'_0}{B_0^2}\left(\frac{l'}{L'_0}
-\frac{2b}{B_0}\right)+\frac{L_0A_0A_{0\theta}}{\Delta_0}\left(
\frac{l}{L_0}+\frac{a}{A_0}+\frac{a_\theta}{A_{0\theta}}\right.\right.\\\nonumber
&\left.\left.-\frac{d}{\Delta_0}\right)\right\}-w_3\left(r+\frac{r^2B'_0}{B_0}
\right)-w_4\frac{C_0C'_0}{B_0^2}+W_{40}\frac{C_0C'_0}{B_0^2}
\left(\frac{c}{C_0}+\frac{c'}{C'_0}-\frac{2b}{B_0}\right)-\omega\left[
2y_1\right.\\\label{ups}
&\left.+Y_{10}\left(\frac{3b}{B_0}+\frac{ar^2A_0B_0^2}{\Delta_0}
+\frac{c}{C_0}\right)\right],\\\nonumber
\Phi&=x_{2\theta}+v'_2+\frac{A_0A'_0}{B_0}\left(\frac{a'}{A'_0}
+\frac{a}{A_0}-\frac{2b}{B_0}\right)V_{10}+\frac{A_0A'_0v_1}{B_0}
+\left(\frac{2B'_0}{B_0}+\frac{B_0^2r^2A_0A'_0}{\Delta_0}\right.\\\nonumber
&\left.+\frac{C'_0}{C_0}+\frac{2rA_0^2B_0^2}{\Delta_0}+\frac{2r^2
A_0^2B_0B'_0}{\Delta_0}\right)v_2+V_{20}\left\{2\left(\frac{b}{B_0}
\right)'+\frac{A_0B_0^2r^2A'_0}{\Delta_0}\left(\frac{a}{A_0}+\frac{a'}
{A'_0}\right.\right.\\\nonumber
&\left.\left.+\frac{2b}{B_0}-\frac{d}{\Delta_0}\right)+\frac{2rA_0^2
B_0^2}{\Delta_0}\left(\frac{2b}{B_0}+\frac{2a}{A_0}-\frac{d}{\Delta_0}\right)
+\frac{2rA_0^2B_0^2}{\Delta_0}\left(\frac{2a}{A_0}+\frac{b'}{B'_0}
+\frac{b}{B_0}-\frac{d}{\Delta_0}\right)\right.\\\nonumber
&\left.+\left(\frac{c}{C_0}\right)'\right\}+x_2\left(
\frac{3B_{0\theta}}{B_0}+\frac{rB_0^2A_0A_{0\theta}}{\Delta_0}
+\frac{r^2A_0^2B_0B_{0\theta}}{\Delta_0}+\frac{C_{0\theta}}{C_0}\right)
+X_{20}\left\{3\left(\frac{b}{B_0}\right)_\theta\right.\\\nonumber
&\left.+\frac{r^2A_0A_{0
\theta}B_0^2}{\Delta_0}\left(\frac{a}{A_0}+\frac{a_\theta}{A_{0
\theta}}+\frac{2b}{B_0}-\frac{d}{\Delta_0}\right)+\left(\frac{c}{C_0}
\right)_\theta+\frac{r^2A_0^2B_0B_{0\theta}}{\Delta_0}\left(
\frac{2a}{A_0}+\frac{b_\theta}{B_\theta}\right.\right.\\\nonumber
&\left.\left.+\frac{b}{B_0}-\frac{d}{\Delta_0}\right)\right\}
-v_3\left(r+\frac{r^2B'_0}{B_0}\right)-r^2v_3\left(\frac{b}{B_0}\right)'
-v_4\frac{C_0C'_0}{B_0^2}-\frac{C_0C'_0}{B_0^2}V_{40}
\left(\frac{c'}{C'_0}\right.\\\label{phi}
&\left.+\frac{c}{C_0}-\frac{2b}{B_0}\right)-{\omega}X_{10}
\left[1+\left(\frac{3b}{B_0}+\frac{ar^2A_0B_0^2}{\Delta_0}
+\frac{c}{C_0}\right)\right].
\end{align}

\end{document}